\documentclass[preprint]{aastex}

\shorttitle{M_{bh}-M_{bulge} relation}
\shortauthors{H\"aring and Rix}

\begin{document}

\title{On the Black Hole Mass - Bulge Mass Relation}

\author{Nadine H\"aring and Hans-Walter Rix}
\affil{Max-Planck Institute for Astronomy, Heidelberg}
\email{haering@mpia-hd.mpg.de and rix@mpia-hd.mpg.de}

\begin{abstract}
We have re-examined the relation between the mass of the central black holes
in nearby galaxies, $M_{bh}$, and
the stellar mass of the surrounding spheroid or bulge, $M_{bulge}$. 
For a total of 30 galaxies bulge masses were derived through Jeans equation
modeling or adopted from dynamical models in the literature. In stellar
mass-to-light ratios the spheroids and bulges span a range of a factor of eight. The bulge masses were related to 
well-determined black hole masses taken from the literature.
With these improved values for $M_{bh}$, compared to \cite{mag98}, and our redetermination of
$M_{bulge}$, we find the $M_{bh}-M_{bulge}$ relation becomes very
tight. We find $M_{bh} \sim M_{bulge}^{1.12\pm0.06}$ with an observed scatter of $\lesssim$ 0.30 dex, 
a fraction of which can be attributed to measurement errors. The scatter in this relation is 
therefore comparable to the scatter in the relations of $M_{bh}$ with
$\sigma$ and the stellar concentration. These results confirm and refine the work of \cite{mar03}.
For $M_{bulge}\sim 5\times 10^{10}
M_{\odot}$ the median black hole mass is $0.14\% \pm 0.04 \%$ of the bulge mass.
\end{abstract}


\section{Introduction}

The good correlations between the mass of the central black
hole and the physical properties of the surrounding stellar bulge provides
evidence that black holes play a key role in the
evolution of galaxies. So far, the tightest relation is that
between the black hole mass $M_{bh}$ and the stellar velocity dispersion
$\sigma$ of the bulge stars \citep{fer00, geb00}. 
Apart from that, other properties correlate with the mass of the
black hole at the center of galaxies. \cite{gra02} showed that $M_{bh}$
correlates tightly with the concentration of the host bulge as quantified by the
Sersic index $n$. 
\cite{mag98} explored the relation between $M_{bh}$ and bulge mass,
$M_{bulge}$, finding $M_{bh}\sim 0.005 M_{bulge}$, but with very
large scatter. It is timely to reconsider this relation, since black hole
mass measurements, now mostly based on HST data, have become much more
reliable. In fact it is now clear that the black hole masses modeled by \cite{mag98} were
overestimated by as much as a factor of ten, since the black hole's sphere of
gravitational influence was not well resolved in their data. This necessarily
implies that the black hole mass fraction is lower than originally estimated \citep{mer01}.
\\
Recently, \cite{mar03} showed that in the near infrared the correlation of the 
bulge luminosity and the black hole mass becomes much tighter than in the
optical. They also demonstrated a tight relation between $R_e \sigma^2_e$ and
the black hole masses, where $R_e \sigma^2_e$ represents a simple virial bulge
mass estimate.\\  
In this Letter we explore further the connection between the mass of the central black
hole and the dynamical mass of the galaxy's bulge or spheroid in more detail,
by combining direct $M_{bh}$ estimates, deemed reliable from other work, with
$M_{bulge}$ determinations based on Jeans equation modeling, as opposed to virial estimates.

\section{The sample}

Our sample consists of 30 nearby galaxies, mostly early types, with existing
reliable black hole mass estimates, of which 27 were drawn from \cite{tre02}. In
addition, the black hole mass for NGC4594 is taken from \cite{kor88}, for
NGC7332 from Gebhardt (private communication), and for NGC4374 from \cite{mac01}. 
For the Milky Way the black hole mass is
from \cite{sch02}. For convenience the appropriate references are listed in
Table \ref{tab1}. 
The selection criteria were first the reliability of the black hole mass and
second the availability of either modeled bulge masses or of surface-brightness and
velocity dispersion profiles, needed for the dynamical modeling of the bulge 
masses.\\  
For 12 of the galaxies we adopted the bulge masses from \cite{mag98}, derived
by them through Jeans equation modeling, after checking that our modeling (cf.\S3)
were consistent for these objects. In addition, we took the bulge masses for NGC1023 from \cite{bower01}, 
for NGC3245 from \cite{bar01},
for NGC4342 from \cite{cretton99} and for NGC3384, for NGC4697, for NGC5845 and for NGC7457 from \cite{geb03}.
\\
The Milky Way is a special case, since the black hole mass is by far the most
secure but the uncertainty in the bulge mass (mostly conceptual) is yet quite
high. Different groups state different values for the bulge mass. The value
$1.1\times 10^{10} M_{\odot}$ is taken from \cite{bis97} and is in agreement
with \cite{dwe95}.
\\
Our sample selection should not introduce any significant bias towards a
particular relation between $M_{bh}$ and $M_{bulge}$ or affect the scatter in such a relation.
The galaxy properties are summarized in Table \ref{tab1}, where group 1 denotes the galaxies modeled 
as part of this work, and the bulge masses for group 2 galaxies were adopted from the literature. 
The distances are taken from \cite{ton01} for most of the galaxies; where these are not available, the distance is
determined from the recession velocity, assuming a Hubble constant of 70 km~s$^{-1}$~Mpc$^{-1}$.    

\clearpage

\begin{table}
\caption[]{Summary of galaxy properties}
\tiny
\begin{minipage}[c]{17.cm}
\begin{tabular}[t]{c c c c c c c c c c } \hline \hline

                     & & & & & & & & & \\

 Galaxy & Type & $\rm{M_{bh} [M_{\odot}]}$ & Ref & $\rm{\sigma [km/s]}$
& $\rm{L_{bulge} [L_{\odot}]}$ & $\rm{\Upsilon, Band [M_{\odot}/L_{\odot}]}$ & $\rm{M_{bulge} [M_{\odot}]}$ & Ref & $\rm{dist [Mpc]}$\\[0.5ex]
(1)&(2)& (3)&(4)&(5)&(6)&(7)&(8)&(9)&(10)\\
& & & & & & & \\
\tableline
group 1 & & & & & & & & \\
\tableline
& & & & & & & &  \\
 M87 & E0 & $3.0^{+1.0}_{-1.0}\cdot 10^9 $ & 1  & 375& $2.0\cdot 10^{11} $ &  3.0,I & 6.0 $\cdot 10^{11}$ & 22, 23, 24 & 16.1\\
 NGC1068 & Sb & $1.4^{+1.3}_{-0.7}\cdot 10^7 $ & 2 & 151 & $1.5\cdot 10^{11} $ & 0.15,R &  $2.3\cdot 10^{10} $ & 25, 26, 27 & 15.0\\
 NGC3379 & E1 & $1.0^{+0.6}_{-0.5}\cdot 10^8 $ & 3 & 206& $1.7\cdot 10^{10} $ &  4.0,R &  $6.8\cdot 10^{10} $ & 28, 29 & 10.6\\
 NGC4374 & E1 & $4.3^{+3.2}_{-1.7}\cdot 10^8 $ & 4 & 296& $6.0\cdot 10^{10} $ &  6.0,R &  $3.6\cdot 10^{11} $ & 30, 23, 24 &  18.4\\
 NGC4261 & E2 & $5.2^{+1.0}_{-1.1}\cdot 10^8 $ & 5 & 315& $4.5\cdot 10^{10} $ &  8.0,R &  $3.6\cdot 10^{11} $ & 30, 23, 24 & 31.6\\
 NGC6251 & E2 & $5.3^{+2.0}_{-4.0}\cdot 10^8 $  & 6 & 290& $9.3\cdot 10^{10} $ & 6.0,R &  $5.6\cdot 10^{11} $ & 31, 32 & 106.0\\
 NGC7052 & E4 & $3.3^{+2.3}_{-1.3}\cdot 10^8 $ & 7 & 266& $8.3\cdot 10^{10} $ & 3.5,R &  $2.9\cdot 10^{11} $ & 33, 28, 34 &  58.7\\
 NGC4742 & E4 & $1.4^{+0.4}_{-0.5}\cdot 10^{7} $ & 8 & 90 & $6.2\cdot 10^{9} $ & 1.0,R & $6.2\cdot 10^{9} $ &  28, 35 & 15.5\\
 NGC821 & E4 & $3.7^{+1.7}_{-1.5}\cdot 10^7 $ & 9 & 209& $2.9\cdot 10^{10} $ & 4.5,R &  $1.3\cdot 10^{11} $ & 28, 29 & 24.1\\
 IC1459 & E3 & $2.5^{+0.5}_{-0.4}\cdot 10^9 $ & 10 & 323& $6.9\cdot 10^{10} $ & 4.2,R &  $2.9\cdot 10^{11} $ & 36, 37 & 29.2\\

& & & & & & & & \\
\tableline
group 2& & & & & & & & \\
 \tableline 

                        & & & & & & & &  \\
  
M31 & Sb  & $4.5^{+4.0}_{-2.5}\cdot 10^7 $ & 11 & 160 & $7.3\cdot 10^{9} $ & 5.1,V &  $3.7\cdot 10^{10} $ & 12 & 0.76\\

M32 & E2  & 2.5$^{+0.5}_{-0.5}\cdot 10^6 $ & 13 & 75 & $3.8\cdot 10^{8} $ & 2.1,V &  $8.0\cdot 10^{8} $ & 12 & 0.81\\

NGC1023 & SB0  & $4.4^{+0.5}_{-0.5}\cdot 10^7 $ & 14 & 205 & $1.2\cdot 10^{10} $ & 5.8,V &  $6.9\cdot 10^{10} $ & 14 & 11.4\\

NGC2778 & E2 & $1.4^{+1.6}_{-0.8}\cdot 10^7 $ & 9 & 175 & $1.2\cdot 10^{10} $ & 6.6,V &  $7.6\cdot 10^{10} $ & 12 & 22.9\\

NGC3115 & S0  & $1.0^{+1.0}_{-0.6}\cdot 10^9 $ & 15 & 230 & $1.7\cdot 10^{10} $ & 7.0,V &  $1.2\cdot 10^{11} $ & 12 & 9.7\\

NGC3245 & S0  & $2.1^{+1.0}_{-0.6}\cdot 10^8 $ & 16 & 205 & $1.7\cdot 10^{10} $ & 3.7,R &  $6.8\cdot 10^{10} $ & 16 & 20.9\\

NGC3377 & E5  & $1.0^{+0.9}_{-0.1}\cdot 10^8 $ & 9 & 145 & $6.4\cdot 10^{9} $ & 2.5,V &  $1.6\cdot 10^{10} $ & 12 & 11.2\\

NGC3384 & S0  & $1.6^{+0.1}_{-0.2}\cdot 10^7 $ & 9 & 143 & $7.1\cdot 10^{9} $ & 2.8,V &  $2.0\cdot 10^{10} $ & 9 & 11.6\\

NGC3608 & E2 & $1.9^{+1.0}_{-0.6}\cdot 10^8 $ & 9 & 182 & $1.9\cdot 10^{10} $ & 5.2,V &  $9.7\cdot 10^{10} $ & 12 & 23.0 \\

NGC4291 & E2 & $3.1^{+1.3}_{-1.1}\cdot 10^8 $ & 9 & 242 & $1.9\cdot 10^{10} $ & 6.9,V &  $1.3\cdot 10^{11} $ & 12 & 26.2\\

NGC4342 & S0  & $3.0^{+1.7}_{-1.0}\cdot 10^8 $ & 17 & 225 & $1.9\cdot 10^{9} $ & 6.3,I &  $1.2\cdot 10^{10} $ & 17 & 15.3\\

NGC4473 & E5 & $1.1^{+0.4}_{-1.0}\cdot 10^8 $ & 9 & 190 & $1.8\cdot 10^{10} $ & 5.2,V &  $9.2\cdot 10^{10} $ & 12 & 15.7\\

NGC4564 & E3 & $5.6^{+0.3}_{-0.7}\cdot 10^7 $ & 9 & 162 & $8.1\cdot 10^{9} $ & 5.4,V &  $4.4\cdot 10^{10} $ & 12 & 15.0 \\

NGC4594 & Sa & $1.0^{+1.0}_{-0.7}\cdot 10^9 $ & 18 & 240 & $4.4\cdot 10^{10} $ & 6.1,V &  $2.7\cdot 10^{11} $ & 12 & 9.8\\

NGC4649 & E1 & $2.0^{+0.5}_{-1.0}\cdot 10^9 $ & 9 & 375 & $6.1\cdot 10^{10} $ & 7.9,V &  $4.9\cdot 10^{11} $ & 12 & 16.8\\

NGC4697 & E4  & $1.7^{+0.2}_{-0.1}\cdot 10^8 $ & 9 & 177 & $2.3\cdot 10^{10} $ & 4.7,V &  $1.1\cdot 10^{11} $ & 9 & 11.7\\

NGC5845 & E3  & $2.4^{+0.4}_{-1.4}\cdot 10^8 $ & 9 & 234 & $6.7\cdot 10^{9} $ & 5.5,V &  $3.7\cdot 10^{10} $ & 9 & 25.9\\

NGC7332 & S0 & $1.3^{+0.6}_{-0.5}\cdot 10^{7} $ & 19 & 122 & $7.9\cdot 10^{9} $ & 1.9,V &  $1.5\cdot 10^{10} $ & 12 & 23.0\\

NGC7457 & S0  & $3.5^{+1.1}_{-1.4}\cdot 10^6 $ & 9 &  67 & $2.1\cdot 10^{9} $ & 3.4,V &  $7.0\cdot 10^{9} $ & 9 & 13.2\\

Milky Way & SBbc &  $3.7 ^{+1.5}_{-1.5}\cdot 10^6 $  & 20 & 75  &       &       &   $1.1\cdot10^{10} $     & 21 &     \\

& & & & & & & & \\

 \tableline
\end{tabular}

\tiny
{
\textsc{Notes}. - (1) Galaxy name. (2) Morphological
  type from \cite{dev91}. (3) Black hole mass taken from \cite{tre02}. (4) References for the black hole masses. (5) Stellar
  velocity dispersions from \cite{tre02}. (6) Bulge luminosity modeled as part of this
  work (group 1) and adapted from the literature (group 2). (7) Mass-to-light
  ratio derived from the Jeans modeling (group 1) and adapted from
  the literature. (8) Modeled bulge mass (group 1) and adapted form the literature (group 2). (9) References for the photometric 
  and kinematic data used for the modeling (group 1) and for the adopted bulge masses (group 2).
  (10) Galaxy distance.\\

\textsc{References}. - (1) \citealt{for94}, (2) \citealt{gre97}, (3) \citealt{geb00b}, (4)
\citealt{mac01}, (5) \citealt{fer96}, (6) \citealt{fer99}, (7) \citealt{vdm98}, (8)
\citealt{tre02}, (9) \citealt{geb03}, (10) \citealt{cap02}, (11) \citet{tremaine95}, (12) \citealt{mag98}, 
(13) \citealt{verolme02}, (14) \citealt{bower01} , (15) \citealt{kormendy96}, (16) \citealt{bar01}, 
(17) \citealt{cretton99}, (18) \citealt{kor88}, (19) Gebhardt (priv. comm.), (20) \citealt{sch02}, (21) \citealt{bis97}\\ 

\textsc{Photometric and Spectroscopic data used for the modeling} (22) \citealt{lau92}, (23) \citealt{pel90}, (24) \citealt{dav88},
 (25) \citealt{san00}, (26) \citealt{pel99}, (27) \citealt{dre84},  (28) \citealt{lau85}, (29) \citealt{ben94}, (30) \citealt{fer94}, 
 (31) \citealt{fer99}, (32) \citealt{hec85}, (33) \citealt{vdb95}, (34) \citealt{din95}, (35) \citealt{dav83}, (36) \citealt{fra89}, 
 (37) \citealt{fri94}

\label{tab1}
}
\end{minipage}
\end{table}
\section{The dynamical modeling}
\begin{figure}
\plottwo{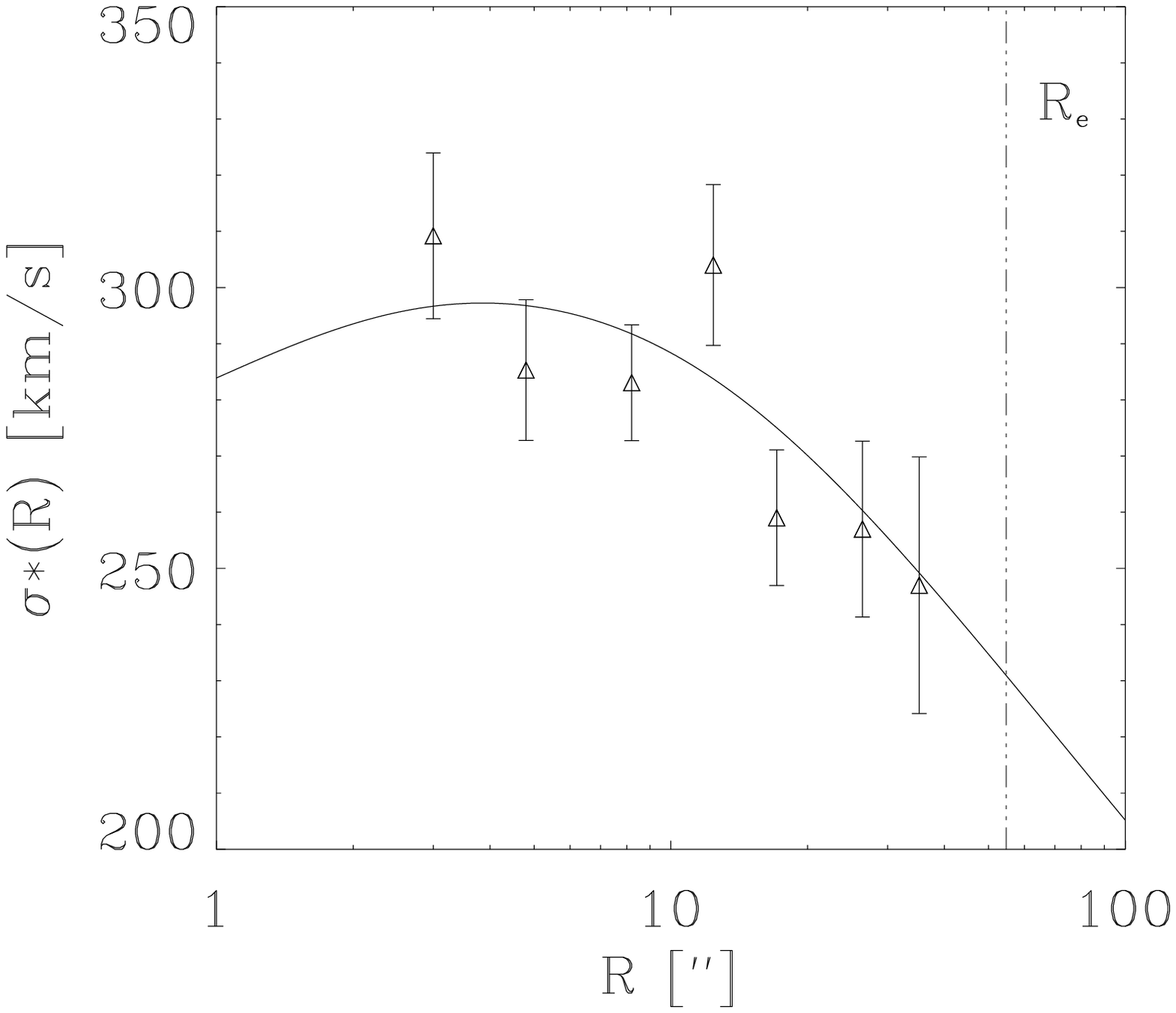}{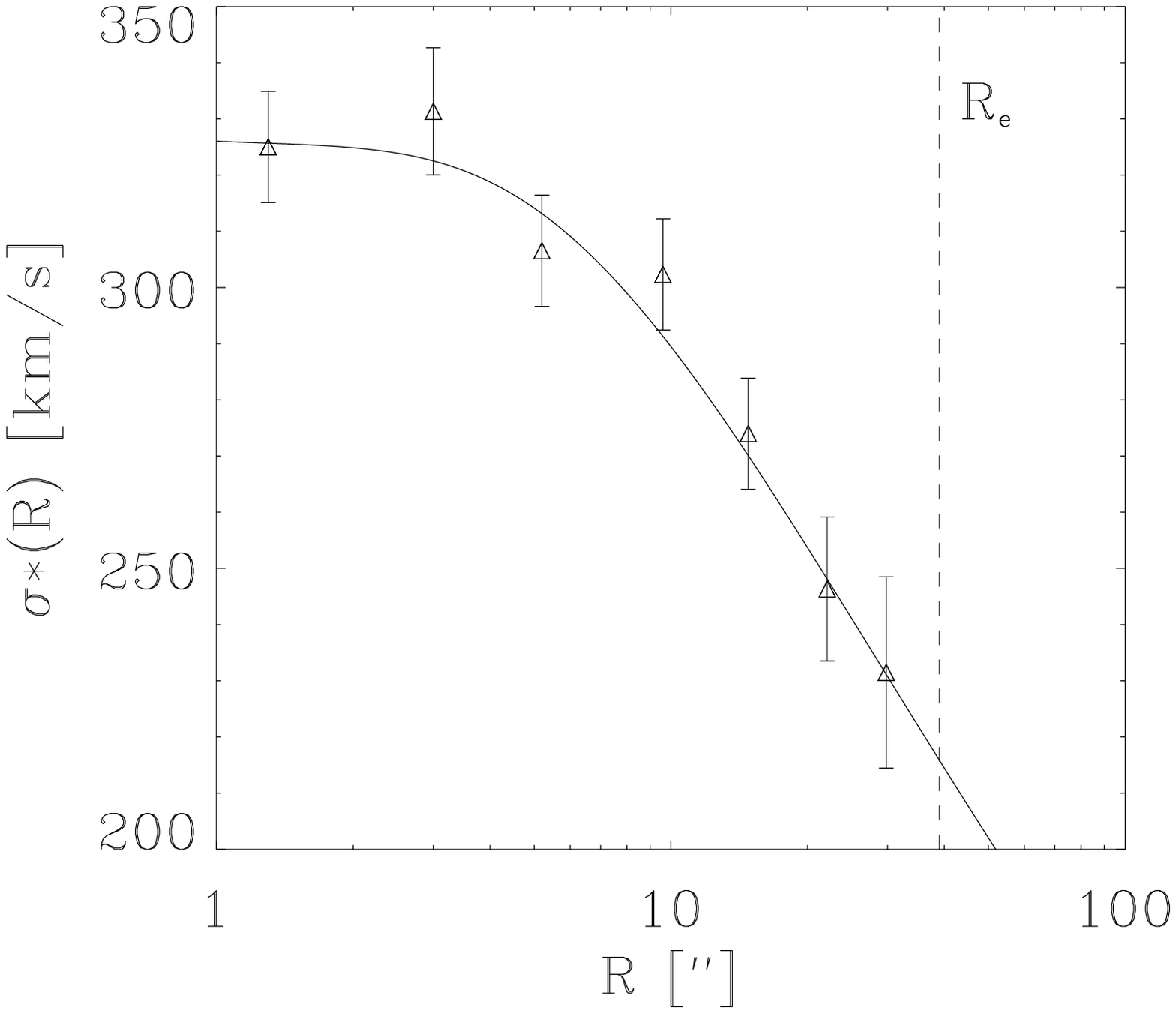}
\caption{Two typical examples for the modeling are shown. The solid line
  gives the modeled velocity profile for NGC4374 (left panel) and NGC4261
  (right panel). The quantity $\sigma^{\star}$ denotes the rms velocity of the stars
  ($\sqrt{\sigma^2+v^2}$). Over-plotted are the data points observed by
  \cite{dav88}. The effective radius, $R_e$, is indicated by the dashed
  line. The modeled velocity profiles match the data well within
  the observed error bars. \label{fig1}} 
\end{figure}

To model the bulge masses of the galaxies in the sample we chose the most
straightforward approach that improves on $M_{bulge}$ estimates from
luminosities or from virial estimates, yet reflects in its simplicity the
inhomogeneous and often scarce kinematic data at larger radii. Specifically,
we solved the Jeans equation in its spherical form:   
\begin{equation}
\frac{d(\rho_{\star} \sigma_r^2)}{dr} + 2\, \frac{\beta \rho_{\star} \sigma^2_r}{r}\,
= -\rho_{\star} \, \frac{d\Phi_{\star}}{dr},
\end{equation}
where $r$ is the radius, $\rho_{\star}$ is the stellar mass density of the
bulge, $\sigma_r$ is the stellar radial velocity dispersion, $\beta$ measures
the anisotropy in the velocity, and $\Phi_{\star}$ is the total potential due to
the stars. In the radial range we cover ($ 1'' \lesssim r \lesssim 30''$), we
neglect any explicit contribution from the dark matter halo. This modeling
procedure is similar to the technique described by \cite{vdM94}. We assumed
the galaxies to be isotropic ($\beta = 0$) and spherically symmetric, which
might lead to an overestimation of the mass. But as \cite{koc94} showed, this
leads to a peak error of not more than 5\% for the velocities. We verified this 
approach by comparison with the axisymmetric models of \cite{mag98}. As boundary
conditions for the modeling we set both the velocity and its first derivative
to vanish at the outer edge of the bulge, which we assume to have a finite 
mass. Since the boundary conditions are set at the outer border of the system,
solving Jeans' equation from the outside to the inside is the appropriate
choice.\\ The procedure is as follows:\\
 1) We fit a surface brightness model to the published photometry, modeling
   the luminosity density as a broken power law, where the inner and outer
   slopes can be fit independently: 
\begin{equation}\label{profile}
 j(r) =
 j_0\,\left(r/a\right)^{-\alpha}\,\left(1+\left(r/a\right)^2\right)^{-\beta},
\end{equation}
with $r$ taken along the major axis of the bulge. The typical radial
range for the fitted profiles extends from $1''$ to $25''$ . 
For these large apertures a seeing correction is not
necessary. All systems, except NGC1068, are bulge dominated and a
bulge-to-disk decomposition is not critical. For NGC1068 we perform a one
dimensional bulge-to-disk decomposition to only account for the bulge stars.\\
 2) A constant mass-to-light ratio $\Upsilon$ is assumed to convert the
  luminosity density into the mass density and calculate the potential.\\
 3) The Jeans equation in its spherical and isotropic form is solved using a
  fourth order Runge-Kutta algorithm, predicting the velocity dispersion
  $\sigma_r(r)$ for each galaxy.\\
 4) The dispersions are integrated along the line-of-sight, projected
  back onto the plane of the sky, averaged over the observational aperture and
  compared to the kinematic data. The observed velocity profiles typically
  extend from $2''$ to $25''$. For these apertures seeing convolution is
  neglected.\\
 5) From this the value for $\Upsilon$ is adjusted by scaling the model velocity
  dispersion curve to best match the observed values. The mass of the
  central black hole is unchanged during the scaling procedure. 
  Two examples of this modeling  procedure are shown in Figure \ref{fig1}.\\
 6) Using this final value for $\Upsilon$, the mass density is integrated over
   the galaxy's bulge (with $r_{max}= 3 r_{\mathit{eff}}$, where the mass of 
   the bulge has already converged).\\
 7) To account for the flattening of the bulge the bulge mass is multiplied by
   $\sqrt{1-\epsilon}$, where $\epsilon$ is the projected ellipticity \citep[e.g.][]{koc94}.\\

The resulting bulge masses determined in this paper are listed in
Table \ref{tab1} (group1), augmented by bulge masses taken from the literature
(group2). We re-modeled three bulge masses from the sample of \cite{mag98} and
found good agreement:\\
M$_{\textrm{\footnotesize M98}}$/M$_{\textrm{\footnotesize here}}$(M87)= 1.3, 
M$_{\textrm{\footnotesize M98}}$/M$_{\textrm{\footnotesize here}}$(NGC821)= 1.0, and
M$_{\textrm{\footnotesize M98}}$/M$_{\textrm{\footnotesize here}}$(NGC3379)= 1.03.

Also in other objects, the mass to light ratio from our spherical Jeans models agree well with
the ones from much more extensive modeling: for IC1459 \cite{cap02} give $\Upsilon_I$= 3.2. 
At a mean color of (R-I) $\sim$ 1/2 (V-I) $\sim$ 0.65 this corresponds to $\Upsilon_R$= 4.1, 
well in agreement with our value,  $\Upsilon_R$= 4.2.
For M87 \cite{vdM94} found $\Upsilon_I$=2.9 in good agreement with our value, $\Upsilon_I$=3.0.
This comparison with a number of other authors show that our models, though more simplistic than other
current approaches, provide sufficiently robust and unbiased estimates of M$_{bulge}$.

To account for the uncertainties introduced by the simplifying assumptions in
the modeling (eg. spherical symmetry, isotropy, constant mass-to-light
ratio) and the inhomogeneity of the data, we give a conservative individual
error estimate for the bulge masses of a factor of 1.5 or $\pm 0.18$dex.  

\begin{figure}[h]
\plotone{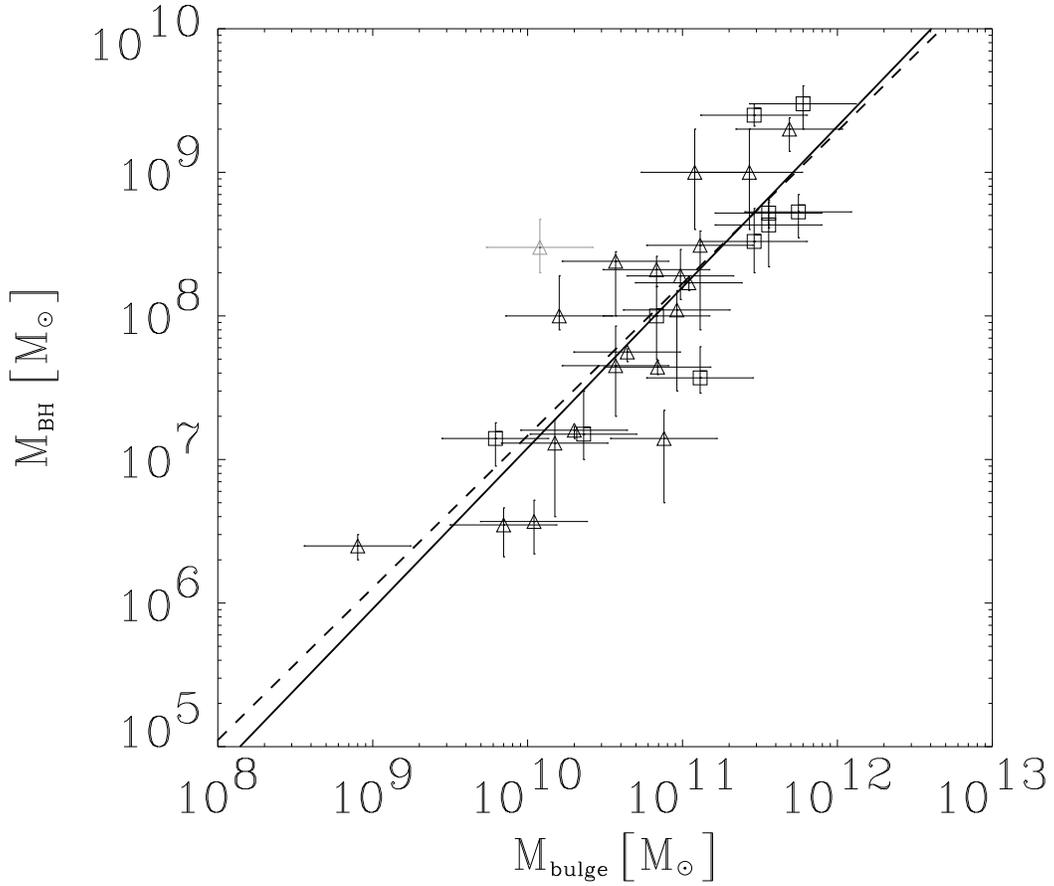}
\caption{Black hole mass vs. bulge mass for the 30 sample
  galaxies. The solid line gives the bisector linear regression fit (see \S4)
  to the data with a slope
  of $1.12\pm0.06$. For comparison the relation found by \cite{mar03} is shown as
  the dashed line (slope: $1.06\pm0.09$). The squares indicate galaxies
  taken from group 1 in Table \ref{tab1} the triangles refer to group 2
  galaxies\label{fig2}. The error bars in black hole mass are the published
  ones given in Table \ref{tab1} and for the bulge mass 
  were adopted to be 0.18 dex in $\log$(M) for all objects. The possible outlier, plotted in 
  light grey is NGC4342 which was not included in the fit.}
\end{figure}
\clearpage

\section{Results and Discussion}
In Figure \ref{fig2}, we plot $M_{bh}$ against the dynamical bulge mass
$M_{bulge}$ for the 30 galaxies in the sample. In contrast to the
relation presented by \cite{mag98}, there is a tight relation without strong outliers. 

A bisector linear regression fit \citep{akr96} to the data leads to the relation 
$$\log(M_{bh}/M_{\odot}) = (8.20\pm0.10)\,+\, (1.12\pm0.06)\,\log(M_{bulge}/10^{11}M_{\odot}),$$\\
where we included constant fractional errors of 0.18 dex on the bulge mass, the published uncertainties for the black
hole masses (see Table \ref{tab1}), and an intrinsic scatter of 0.3 dex.
For this fit we have excluded the one possible outlier, NGC 4342; its inclusion leaves 
the slope of the relation unchanged. 
We calculated the mean values and their 1$\sigma$ uncertainties using the bootstrap
method \citep{efr93}. Bootstrapping is preferable, since we do not have
rigorous error bars for $M_{bulge}$, or for all $M_{bh}$ estimates. 
Whether we adopt intrinsic scatter or $\delta$M$_{bulge}$ of 0.2 dex or 0.3
dex for the fit changes the slope by $\lesssim$1\% and the intercept by less than 0.3\%. A
least squares fit using the FITEXY routine \citep{press92} changes the slope
by $\sim$15\%, yielding 1.32$\pm0.17$. 
All values agree within the stated uncertainties.
\\ 
The upper limit on the intrinsic dispersion in the $M_{bh}-M_{bulge}$
relation, namely the observed dispersion assuming no measurement errors, 
is $\sim 0.30$ dex. A significant portion of this scatter can plausibly be attributed to the
observational errors in black hole masses. 
The implied median black hole mass fraction at bulge masses of $\sim 5\times 10^{10}M_{\odot}$ is
$M_{bh}$/M$_{bulge}=(1.4\pm0.4)\cdot10^{-3}$. This fraction is in
agreement with the estimates from \cite{mer01} and \cite{mar03}.
\\ 
At face value, the slope of the $M_{bh}-M_{bulge}$ relation exceeds unity with
1.5$\sigma$ significance, both for the Akritas \& Bershady estimator and the 
FITEXY routine, as also used by \cite{mar03}. However, the
data are still in agreement with a $M_{bh}-M_{bulge}$ proportionality at the $< 2 \sigma$ level
and we do not want to emphasize the non-linearity. 

The mass-to-light ratios $\Upsilon$ we found through 
the dynamical models are spread over a wide range from 0.15 to 8.0 
$M_{\odot}/L_{\odot}$. Excluding the smallest value, which comes 
from NGC1068 (a galaxy with starburst activity), we still 
find a range for $\Upsilon$ of a factor of eight. 
\\
Revisiting the Magorrian relation with more reliable black 
hole masses leads to a relation with a strongly reduced scatter. 
Our relation shows at most one outlier (NGC4342, among 30 objects) 
and we did not apply any selection criteria apart from the reliability of 
the black hole masses. 
\\ 

Our results confirm and expand the findings of
\cite{mar03}, who relate black hole masses to infrared
luminosities and also to virial bulge mass estimates ($M\sim \sigma^2
r_e$). Their $M_{bh}-M_{bulge}$ relation for all galaxies is statistically in
agreement with the relation we find. They find a slightly higher observed
scatter, potentially because their virial estimate is less precise than the
Jeans equation estimates, used here.
\\ 
Through determining and compiling M$_{bulge}$ measurements for the objects with robust M$_{bh}$ estimates,
we could demonstrate that the scatter in the black hole mass to bulge mass relation is nearly as small as
that in the $M_{bh}-\sigma$ \citep{geb00,fer00} ($\sim$0.3 dex) and  
$M_{bh}$-concentration \citep{gra02} ($\sim$0.31 dex) relation. Therefore, the 
relation between the black hole mass and the velocity dispersion 
is not unique and it seems as if the large scatter in the original 
Magorrian relation is due to erroneous estimation of the black 
hole masses. 
\\
Still, in the local universe $M-\sigma$ is of invaluable
practical use, since velocity dispersions are easy to measure. 
However, towards higher redshift ($z\gtrsim 2$) the relation between black
hole mass and stellar bulge mass gains importance. It is then exceedingly
difficult to measure the velocity dispersion, but the bulge mass can be
estimated via the measured luminosity and an upper limit of the stellar
mass-to-light ratio, derived from the maximal age of the stellar population at
that redshift.

\acknowledgments
We thank Andi Burkert for stimulating discussions on this paper, and Tim de Zeeuw and 
Michele Cappellari for useful comments. We thank the anonymous referee for a
thorough and constructive report that helped to improve the manuscript.

\clearpage
\end{document}